\documentclass[11pt,twoside]{article}
\usepackage{asp2010}

\resetcounters
\begin{document}

\title{Beads on a String and Spurs in Galactic Disks}
\author{F.~Renaud,$^1$ F.~Bournaud,$^1$ E.~Emsellem,$^{2,3}$ B.~Elmegreen,$^4$ and R.~Teyssier$^5$
\affil{$^1$Laboratoire AIM Paris-Saclay, CEA/IRFU/SAp, Universit\'e Paris Diderot, F-91191 Gif-sur-Yvette Cedex, France}
\affil{$^2$European Southern Observatory, 85748 Garching bei Muenchen, Germany}
\affil{$^3$Universit\'e Lyon 1, Observatoire de Lyon, CRAL et ENS, 9 Av Charles Andr\'e, F-69230 Saint-Genis Laval, France}
\affil{$^4$IBM T. J. Watson Research Center, 1101 Kitchawan Road, Yorktown Heights, New York 10598 USA}
\affil{$^5$Institute for Theoretical Physics, University of Z\"urich, CH-8057 Z\"urich, Switzerland}}


\begin{abstract}
The organization of the interstellar medium in disk galaxies obeys the large scale dynamics set by kpc-size structures. Improving our knowledge of how the dense, molecular gas is distributed in a disk is an important step in our understanding of star formation at galactic scale. Using a recently published simulation of a Milky Way-like galaxy, we explore the formation and dynamical organization of the star forming gas in a proto-typical disk. Along spiral arms, we report the formation of regularly spaced clouds, called beads on a string and spurs. The former form through gravitational instabilities while the later originate from Kelvin-Helmholtz instabilities. We propose that the co-existence of both structures in the same galaxy can be explained by a different role of the disk dynamics, depending on the location within the disk. In particular, we highlight the impact of the pitch angle of the spiral arm in the development of either type of structure.
\end{abstract}

\section{Introduction}

In disks, number of structures are governed by and govern the dynamics of galaxies. Together with bars, spiral arms host most of the star formation in a disk \citep[see][and many references since then]{Morgan1953}. Spiral waves are often seen as shocks increasing the local density of gas and triggering its collapse into clouds, cores and stars \citep{Elmegreen1979, Dobbs2009}. The details of the formation and the nature of these clouds are therefore of prime importance for star formation in disks. In this contribution, we summarize the properties of two major families of clouds, the beads on a string and the spurs, and propose an explanation for their different origin.

\section{A quick sensus of structures}

\begin{figure}
\begin{center}
\includegraphics{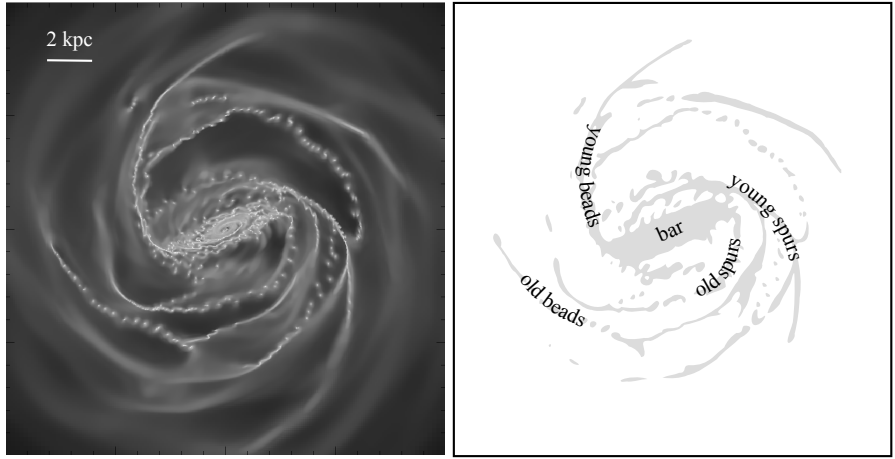}
\caption{Surface density of the gas disk, and locations of the main groups of beads on a string and spurs. Beads retain the gas material along the spiral arm, while spurs are off-centered, on the leading side the arm.}
\label{fig:map}
\end{center}
\end{figure}

Simulations dedicated to the study of the interstellar medium (ISM) of galaxies often set a background potential, usually made of a spiral pattern and/or a bar, and follow the evolution of a ``live'' gaseous component \citep[e.g.][]{Bonnell2006, Dobbs2006, Tasker2009}. Such well-defined configurations ease the physical interpretation, but miss the diversity of structures that disks host. \citet{Renaud2013b} recently presented a self-consistent hydrodynamical simulation of a Milky Way-like galaxy, and studied the large-scale structure of the ISM. The complexity of the stellar component and the gas respond to its potential create a diversity of structures along the spiral arms, as shown in Figure~\ref{fig:map}\footnote{An interactive map is available online: \url{http://irfu.cea.fr/Pisp/florent.renaud/mw.php}}.

In this model of the Milky Way, we note the presence of beads on a string, similar to those observed in several nearby galaxies \citep[e.g.][]{Elmegreen1983, Foyle2013}. These regularly spaced clouds are formed by the gravitational collapse of their host arm, within $\sim 10$ Myr. The clouds further grow by accreting surrounding gas, which progressively increases the cloud / inter-cloud density contrast and thus empties the gas content of the ``string'' connecting the ``beads''. Elsewhere in the disk, other spirals host spurs instead of beads. Kelvin-Helmholtz (KH) instabilities were the first explanation given for the formation of spurs \citep{Wada2004}, but other works questioned this conclusion, invoking an over-idealization of these models (e.g. the lack of magneto-hydrodynamics or the 2D aspect, see \citealt{Kim2006, Shetty2006, Dobbs2006} and \citealt{Renaud2013b} for a short review). Their presence in our simulation demonstrates that spurs can be formed in a non-idealized setup, without magnetic fields, and with parameters in reasonable agreement with observations.

The main morphological difference between beads and spurs is their position with respect to the spiral wave. Beads on a string form \emph{in situ} via gravitational instability around dense seeds along the spiral arm. On the contrary, spurs are made of material somewhat ejected from the bulk of the spiral by KH instabilities. We found in our simulation that spurs remain off the spiral for at least several $10^7$ Myr instead of e.g. falling back onto the arm, which is supported by the observations of \citet{Elmegreen1980} who suggested that the spurs are as long-lived as the spirals. Once gas is sent on the leading side of the spiral, it condenses at the tip of the spurs, becomes self-gravitating, and forms star clusters. Since the spurs are offset with respect to the spiral, star formation occurs before the passage of the arm, i.e. without the ``trigger'' of a density wave or a shock. Such offset star formation is visible in the recent observations of M~51 by \citet[see e.g. their Figure 8]{Schinnerer2013}. 

\section{Role of the pitch angle}

\begin{figure}
\begin{center}
\includegraphics{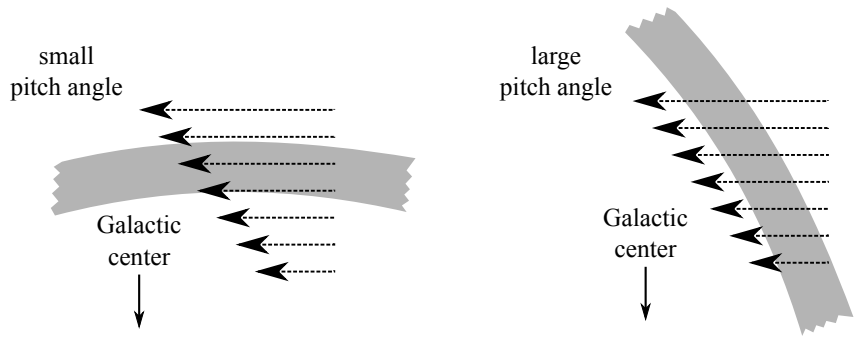}
\caption{Sketch of a spiral arm (grey shade) in the case of a small pitch angle (left) and a large one (right). The Galactic velocity field, increasing with radius, is represented with dashed arrows. The velocity gradient across a spiral arm depends on the pitch angle. Small pitch angles would favor the formation of spurs through KH instabilities.}
\label{fig:pitch}
\end{center}
\end{figure}

Figure~\ref{fig:pitch} shows that in the case of a small pitch angle, the velocity gradient across the arm is higher than for larger pitch angles. Such difference could lead to a different response of the gaseous material of the spiral to KH instabilities and thus, as we stated above, to the formation of spurs. Rigorously speaking, the important quantity would be the angle between the spiral and the \emph{local} velocity vector, but to the first order, one can reasonably assume that the velocity vector is tangential to the galactic radius, and thus that this quantity is directly related to the pitch angle.

Therefore, for a given radial gradient of the velocity field, the organization of the ISM into spurs or beads on a string would depend on the pitch angle. We note however that the pitch angle of a given portion of the spiral changes along the evolution of the galaxy. Only the angle at the formation epoch of the seeds of gas clumps would matter in determining the formation of spurs or beads, since self-gravity of the clumps should take over in a few Myr and dominate the further evolution of gas packets. \citet{Wada2004} introduced a mathematical formalism providing a condition for the KH instability in term of Richardson number ($< 1/4$), following \citet{Chandrasekhar1961}, and reached the same conclusions than the qualitative argument given here \citep[see also][]{Lee2012}.

\citet{Elmegreen1980} studied the separation and angle of spurs with respect to the spiral\footnote{The pitch angle defined in \citet{Elmegreen1980} represents ``the angle between the spur and a circle ar the same radius as the starting point of the spur'', which is not the same as the pitch angle of spiral used below.} and found relatively universal values ($\sim 500$ pc and $\sim 60 \deg$). Our simulation also shows regular separation and angle along a given arm, but these values seems to vary from one arm to the next, because of an evolutionary process. We measure that the spurs in the structure labelled ``old spurs'' in Figure~\ref{fig:map} separate from each other at $\sim 15$ km/s, because of differential rotation. Since large pitch angles means high differential rotation along an arm, the separation between the spurs would increase in arms with large pitch angles. We note that, on Gyr-long time-scales, the changes of the spiral pattern are likely to drive a more complex evolution of the separation of the spurs than the one we picture here, if these structures are indeed long-lived.

In conclusion, the pitch angle angle of a spiral arm is connected to the nature of the star forming clouds (beads or spurs), at the time of their formation. Later, the separation between the clouds can also be connected to the pitch angle, via differential rotation.

\acknowledgements FR thanks Marc Seigar and the SOC/LOC for organizing a very interesting and enjoyable conference.

\end{document}